\documentclass[prb,twocolumn,showpacs]{revtex4}

\usepackage[dvips]{graphicx}

\begin{document}

\title{Doping Dependence of Anisotropic Resistivities in Trilayered Superconductor Bi$_2$Sr$_2$Ca$_2$Cu$_3$O$_{10+\delta}$ (Bi-2223)}

\author{Takenori Fujii}
 \altaffiliation[Corresponding author, Present address: ]{Department of Applied Physics, School of Science and Engineering, Waseda University, Tokyo 169-8555, Japan}
 \email{fujii@htsc.phys.waseda.ac.jp}
\affiliation{Department of Applied Physics, Faculty of Science, Tokyo University of Science, Tokyo 162-8601, Japan}

\author{Ichiro Terasaki}
\affiliation{Department of Applied Physics, School of Science and Engineering, Waseda University, Tokyo 169-8555, Japan}

\author{Takao Watanabe}
 \altaffiliation[Present address: ]{NTT Photonics Laboratories, Kanagawa 243-0198, JAPAN.}
\affiliation{NTT Basic Research Laboratories, Kanagawa 243-0198, Japan}

\author{Azusa Matsuda}
\affiliation{Department of Applied Physics, Faculty of Science, Tokyo University of Science, Tokyo 162-8601, Japan\\NTT Basic Research Laboratories, Kanagawa 243-0198, Japan}

\date{\today}

\begin{abstract}
The doping dependence of the themopower, in-plane resistivity $\rho_{ab}$($T$), out-of-plane resistivity $\rho_c$($T$), and susceptibility has been systematically measured for high-quality single crystal Bi$_2$Sr$_2$Ca$_2$Cu$_3$O$_{10+\delta}$. We found that the transition temperature $T_c$ and pseudogap formation temperature $T_{\rho_c}^*$, below which $\rho_c$ shows a typical upturn, do not change from their optimum values in the "overdoped" region, even though doping actually proceeds. This suggests that, in overdoped region, the bulk $T_c$ is determined by the always underdoped inner plane, which have a large superconducting gap, while the carriers are mostly doped in the outer planes, which have a large phase stiffness.
\end{abstract}

\pacs{74.25.Fy, 74.62.-c, 74.72.Hs}

\maketitle

The Bi-Sr-Ca-Cu-O system consists of many superconducting phases with a number of CuO$_2$ planes in a unit cell. In the bilayer Bi$_2$Sr$_2$CaCu$_2$O$_{8+\delta}$ (Bi-2212) system, the CuO$_2$ planes are homogeneously doped, since these planes are crystallographically equivalent. On the other hand, the trilayered Bi$_2$Sr$_2$Ca$_2$Cu$_3$O$_{10+\delta}$ (Bi-2223) system has two crystallographically inequivalent CuO$_2$ planes, an inner CuO$_2$ plane with a square (four) oxygen coordination and two outer CuO$_2$ planes with a pyramidal (five) oxygen coordination. Then, there is a possibility of inhomogeneous doping amoung layers. Recently, Kivelson~\cite{kivelson} proposed that such a inhomogeneous doping helps to increase $T_c$ and accounts for the higher $T_c$'s in the multi-layered system. There, a high pairing energy scale is derived from the underdoped planes and a large phase stiffness from the optimally or overdoped ones. The combination of these two may provide a key to achieve the higher $T_c$ in the cuprate system. Therefore, it is very important to study the actual muli-layer system in detail.

In high $T_c$ cuprates, there is a consensus that the sets of CuO$_2$ planes separated by the blocking layer are only weakly coupled and the interaction between them can be understood as a tunneling process. This is known as the confinement efect from the theoretical point of view~\cite{PW}. However, it is still unknown whether the confinement works for the CuO$_2$ planes within the unit cell, that was assumed in the above mentioned Kivelson's theory. The inhomogeneous charge distribution provides a new way of investigating such an effect.

The difference of carrier concentration between the inner and outer planes has been reported in NMR studies of multilayered systems~\cite{NMR, NMR2, NMR3, NMR4, NMR5, Ctoku}. In the case of (Cu$_{0.6}$C$_{0.4}$)Ba$_2$Ca$_3$Cu$_4$O$_{12+y}$, which consists of two inner planes and two outer planes, it was reported that the magnetic and superconducting properties are distinctly different between the inner and outer planes, and the bulk $T_c$ is triggered by the underdoped inner planes. However, all these experiment were performed by using polycrystalline sample, which is magnetically alligned along the $c$-axis and there have been few investigations of precise doping dependence of single crystal because high-quality samples of multilayered system have not been available. Here, we have successfully grown high-quality single crystals of the trilayered system Bi-2223~\cite{Cfujii}, and measured the doping dependence of in-plane resistivity $\rho_{ab}$, out-of-plane resistivity $\rho_c$, thermopower $S$, and normal state susceptibility $\chi$ for the first time. Based on the results, we discuss the possibility of charge distribution among the CuO$_2$ planes and the interaction between the CuO$_2$ planes within the unit cell.

High-quality single crystals were grown using the traveling solvent floating zone (TSFZ) method~\cite{Cfujii}. The X-ray diffraction pattern showed only sharp Bi-2223 peaks, confirming the good crystallinity of our samples. The $c$-axis length was estimated from the fitting method using the Nelson-Riley function. The oxygen content $\delta$ was controlled by annealing a sample with varying Ar and O$_2$ gas flow ratios and/or temperatures. A highly oxygenated sample was prepared by high O$_2$ pressure (400 atm) annealing using a hot isostatic pressing (HIP) furnace. The annealing conditions for Bi-2223 samples used in this paper are a: O$_2$ 5$\times$10$^{-3}$ torr 600$^\circ$C, b: O$_2$ 0.01\% 600$^\circ$C, c: O$_2$ 0.1\% 600$^\circ$C, d: O$_2$ 1\% 600$^\circ$C, e: O$_2$ 10\% 600$^\circ$C, f: O$_2$ 600$^\circ$C, g: O$_2$ 500$^\circ$C, h: O$_2$ 400$^\circ$C, i: HIP O$_2$ 400atm 500$^\circ$C, (These discriptions are used in all Figures). The superconducting transition temperatures $T_c$ were defined by the onset of the Meissner effect. For normal state susceptibility measurement, we used a large single crystal ($\approx$10 mg) and applied high magnetic field (5 T). $\rho_{ab}$ was measured with the standard four-probe method, while $\rho_c$ was measured with four-probe-like method with the voltage contacts attached to the center of the $ab$ plane and the current contacts covering almost all of the remaining surface~\cite{wataB}. The thermopower was measured using a steady-state technique, where a temperature gradient of 1 K/cm was generated by a small resistive heater and was monitored by differential thermocouple made of copper-constantan.

\begin{figure}
 \includegraphics[width=7cm,clip]{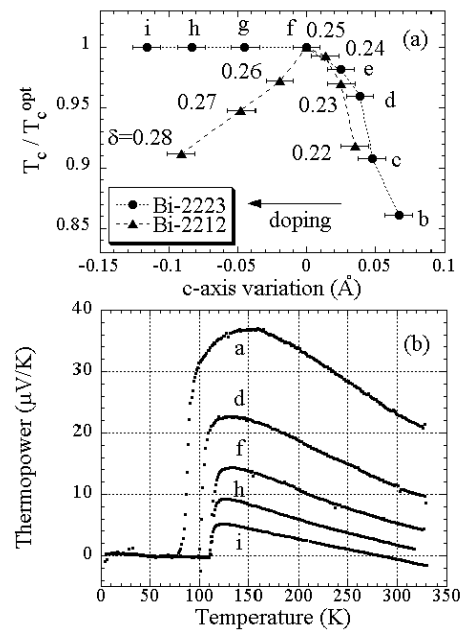}
 \caption{
 (a) Normalized transition temperature $T_c$ plotted against the $c$-axis variation. The $T_c$ and $c$-axis length of optimum-doping samples are 89 K, 30.864 $\AA$\quad for Bi-2212 and 108 K, 37.119 $\AA$\quad for Bi-2223 respectively. The annealing conditions for Bi-2223 are a: O$_2$ 5$\times$10$^{-3}$ torr 600$^\circ$C, b: O$_2$ 0.01\% 600$^\circ$C, c: O$_2$ 0.1\% 600$^\circ$C, d: O$_2$ 1\% 600$^\circ$C, e: O$_2$ 10\% 600$^\circ$C, f: O$_2$ 600$^\circ$C, g: O$_2$ 500$^\circ$C, h: O$_2$ 400$^\circ$C, i: HIP (O$_2$ 400atm 500$^\circ$C), while the $\delta$ of Bi-2212 are determined by the $P_{O_2}-T$ phase diagram obtained by our previous thermogravimetric measurement. (b) Thermopower $S$ of Bi$_2$Sr$_2$Ca$_2$Cu$_3$O$_{10+\delta}$ single crystal annealed in various atmospheres [The labels in Figs. 1, 2, 3, and 4 correspond to each other].
 }
\label{f1}
\end{figure}

Figure \ref{f1}(a) shows the normalized transition temperature $T_c$ plotted against the relative change of the $c$-axis length from that of the sample "f". The relation between the $T_c$ and $c$-axis length of Bi-2212 is also plotted for comparison. (The $\delta$ of Bi-2212 is determined by the $P_{O_2}-T$ phase diagram obtained by our previous thermogravimetric measurement~\cite{wata1}.) The $T_c$ and $c$-axis length are 89 K, 30.864 \AA\quad for optimally-doped Bi-2212 and 108 K, 37.119 \AA\quad for Bi-2223 (sample "f") respectively. The $c$-axis length monotonically decreases with increasing $\delta$ both in Bi-2212 and Bi-2223, indicating that oxygen is actually incorporated into crystals. In the case of Bi-2212, reflecting the bell-shaped doping dependence of $T_c$, which is common behavior in mono- or bilayer cuprates, $T_c$ increases with decreasing $c$-axis length, reaches its maximum, and then decreases with decreasing $c$-axis length. On the other hand, $T_c$ of Bi-2223 increases with decreasing $c$-axis length quite similar to Bi-2212. However, $T_c$ keeps its maximum value~\cite{ihara} when the $c$-axis length is further decreased.

To confirm the carrier doping in the constant $T_c$ region, we measured the doping dependence of the thermopowers $S$ (Fig. \ref{f1}(b)). The magnitude of the thermopower monotonically decreases with increasing $\delta$. This result clearly shows that the carrier was doped continuously even in the constant $T_c$ region. The room temperature thermopower is considered to be an universal measure of the doping level~\cite{obert}. By using this measure, the (average) doping level of sample "f" can be assigned to "optimal" doping. Then, the sample "i" would be assigned to slightly overdoping with carrier concentration about p=0.185. Thus it would show the $T_c$ of 102 K, if the carriers were homogeneously doped. We will call, hereafter, the constant $T_c$ region as overdoped region.

\begin{figure}
 \includegraphics[width=7cm,clip]{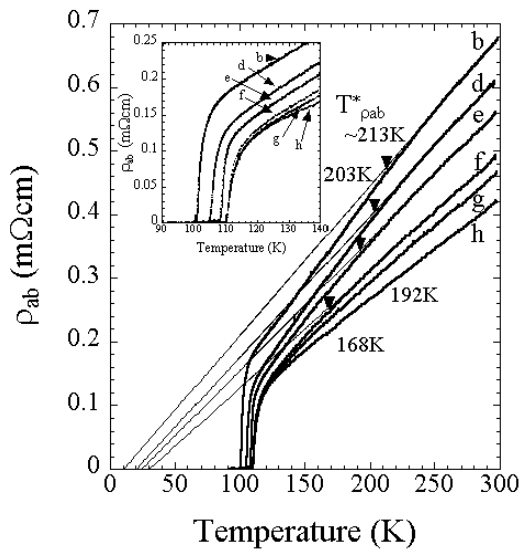}
 \caption{
 In-plane resistivity $\rho_{ab}$ of Bi$_2$Sr$_2$Ca$_2$Cu$_3$O$_{10+\delta}$ single crystal annealed in various atmospheres. The solid straight lines, which are linear extrapolations of $\rho_{ab}$ at higher temperatures, are shown as guidelines. The temperatures $T^*_{\rho_{ab}}$ at which the $\rho_{ab}$ deviates from $T$-linear behavior are shown by arrows. The scale is expanded in the inset for a better view around $T_c$.
 }
\label{f2}
\end{figure}

The temperature dependence of in-plane resistivity $\rho_{ab}(T)$ with various $\delta$ is shown in Fig. \ref{f2}. The absolute values of $\rho_{ab}$ and the overall slopes $d\rho_{ab}$/$dT$ monotonically decrease with increasing $\delta$, indicating that the carriers are actually doped with increasing $\delta$. In all doping level, they show negative residual resistivity as indicated by the solid lines in Fig. \ref{f2}. High-$T_c$ materials with $T_c$ larger than 100 K tend to show negative residual resistivity. The Bi-2223 also seems to belong this class, although we do not know its relevance to $T_c$. As seen in the inset of Fig. \ref{f2}, $T_c$ determined by zero resistivity increases from 100 to 110 K with increasing doping level from "b" to "f". However, in the overdoped region, $T_c$ does not change from 110 K. The underdoped samples (denoted by b, d, e, and f) show a downward deviation from high-temperature $T$-linear behavior below a certain temperature $T^*_{\rho_{ab}}$, similarly to that of Bi-2212~\cite{wata1}. $T^*_{\rho_{ab}}$ increases with decreasing doping as $T^*_{\rho_{ab}}$=168, 192, 203, and 213 K for the samples labeled as f, e, d, and b, respectively as indicated by arrows in Fig. \ref{f2}. Here, $T^*_{\rho_{ab}}$ was determined as a temperature at which $\rho_{ab}$ deviates 1\% from the high temperature $T$-linear resistivity using a similar analysis shown in Ref. \cite{itoPRL}.

\begin{figure}
 \includegraphics[width=7cm,clip]{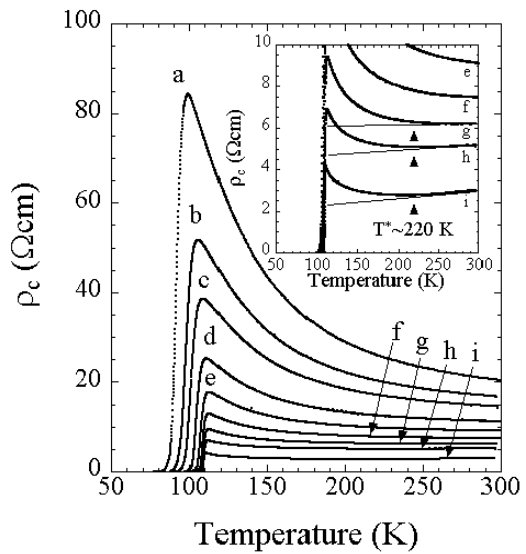}
 \caption{
 Out-of-plane resistivity $\rho_c$ of Bi$_2$Sr$_2$Ca$_2$Cu$_3$O$_{10+\delta}$ single crystal annealed in various atmospheres. Overdoped behavior of out-of-plane resistivity is shown in the inset. The solid straight lines in the inset, which are linear extrapolations of $\rho_c$ at higher temperatures, are eye guides for the overdoped sample. Arrows indicate the temperatures $T^*_{\rho_c}$ below which $\rho_c$ shows a characteristic upturn.
 }
\label{f3}
\end{figure}

Figure \ref{f3} shows the c-axis resistivity $\rho_c(T)$ for various $\delta$. Similarly to $\rho_{ab}$, we can see that $T_c$ is pinned at the maximum value in the overdoped region. The overall magnitude of $\rho_c$ decreases with increasing $\delta$. We have previously shown that the pseudogap formation in the elastic (coherent) tunneling model is an effective explanation for the insulating $\rho_c$~\cite{wata2}. The decrease in the absolute value of $\rho_c$ with $\delta$ would imply an increase in the in-plane density of states (DOS) and semiconductive behavior would be attributed to the decrease of DOS due to the pseudogap formation. The underdoped samples from "a" to "f" show semiconductive $\rho_c$ in all temperature regions measured. As seen in the inset of Fig. \ref{f3}, in samples "g", "h", and "i", $\rho_c$ decreases linearly with decreasing temperature at higher temperature and show a semiconductive upturn below the characteristic temperature $T^*_{\rho_c}$ (shown by the arrow in the inset of Fig. \ref{f3}). Here, the straight lines are linear extrapolations of $\rho_c$ at higher temperatures. We estimated $T_{\rho_c}^*$ as the temperature below which $\rho_c$ deviates 1\% from the linear straight lines~\cite{takenaka}. Recent ARPES experiment showed that the pseudogap begins to open from the ($\pi$, 0) direction (hot spot)~\cite{ARPES, ARPES2}. $\rho_c$ should be particularly sensitive to the onset of the pseudogap formation~\cite{shiba}, since the hopping probability $t_c$ in the $c$-axis direction will be dominated by carriers around hot spot on the anisotropic Fermi surface [It is expressed as $t_c\sim$(cos$k_xa$-cos$k_ya$)$^2$]~\cite{ioffe}. We can see that the $T_{\rho_c}^*$ remains unchanged ($\approx$220 K) for doping levels higher than "g", while the absolute value of $\rho_c$ continues to decrease.

\begin{figure}
 \includegraphics[width=7cm,clip]{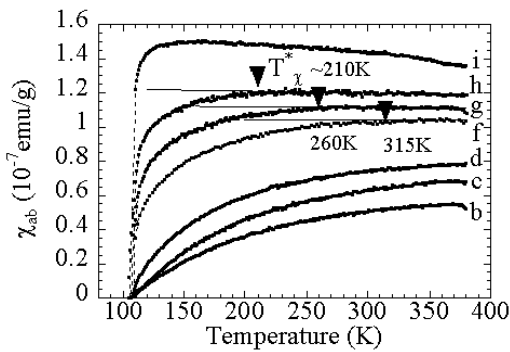}
 \caption{
 Magnetic susceptibilities $\chi_{ab}$ of 
Bi$_2$Sr$_2$Ca$_2$Cu$_3$O$_{10+\delta}$ single crystal annealed 
in various atmospheres. The solid straight lines in the inset, which are linear extrapolations of $\chi_{ab}$ at higher temperatures, are eye guides for the overdoped sample. Arrows indicate the temperatures $T^*_{\chi}$ at which $\chi_{ab}$ start to decrease from its linear high temperature behaviors.
 }
\label{f4}
\end{figure}

The magnetic susceptibilities $\chi_{ab}(T)$ for various $\delta$ are shown in Fig. \ref{f4}, where a magnetic field of 5 T was applied parallel to $a$-axis. The overall magnitude of $\chi_{ab}$ monotonically increases with increasing $\delta$. We interpret this in terms of an increase in the DOS near the Fermi level with carrier doping. At all doping levels, the behavior of temperature dependence of $\chi_{ab}$ is quite similar to that of Bi-2212~\cite{wata2}. The susceptibilities for the underdoped sample (denoted by b, c, and d) monotonically decrease with decreasing temperature, implying a decrease in DOS due to the pseudogap formation. The sample near the optimum-doping level (denoted by f, g, and h) show Pauli paramagnetic behavior at high temperature. And then, the susceptibilities decrease below characteristic temperature $T^*_{\chi}$. $T^*_{\chi}$ is estimated as 210, 260, and 315 K for the sample labeled h, g, and f, respectively. Here, $T^*_{\chi}$ was determined as the temperature at which $\chi_{ab}$ deviates 1\% from high-temperature $T$-linear behavior. Heavily oxygenated (HIPed) sample "i" shows negative temperature dependence ($d\chi/dT \leq$ 0), and the dependence is approximately linear. We consider this behavior to be an anomalous DOS effect due to the existence of a van Hove singularity for an overdoped sample~\cite{wata2,matsuda}. 

The anomalous $T_c$ and $T^*_{\rho_c}$ pinning in the overdoped region can be understood by considering inequivalent hole doping between the inner and outer planes. In the overdoped region, the carriers would be mostly doped in the outer planes, and the inner plane would remain at the underdoping level. Here, the coupled system, which Kievelkson assumed, seems to be realized. Highest Tc's would be maintained by the combination of inner plain's large superconducting gap and the outer plain's large superfluid density~\cite{uemura, uemura2}, which controls the stiffness of the system to phase fluctuations. From the practical point of view, the phase stiffness is very important for sustaining a large superconducting curent, as well as $T_c$. This gives a promissing way of improving the characteristics of high $T_c$ materials.

On the other hand, $T^*_{\rho_c}$ pinning in the overdoped region, as well as semiconducting $\rho_c$ behavior, is also determined by the always optimally-doped inner plane, which is known to have a pseudogap. Since the outer planes are overdoped, the transport between outer planes separated by the Bi$_2$O$_2$ layer may not show a pseudogap effect. Then the observed pseudogap effect may come from the transport between the outer and inner planes. This indicates that the interaction between the CuO$_2$ planes within a unit cell is very weak, like that between the CuO$_2$ planes separated by the blocking layer. Thus, in multilayered system, there already exist an array of weakly coupled planes, which was assumed in the above-mentioned Kivelson's theory. 

In Bi-2212, the characteristic temperatures $T^*_\chi$ and $T^*_{\rho_c}$ coincide for all doping levels and shift to lower temperature with increasing doping level~\cite{wata2}. Thus, as mentioned above, semiconductive $\rho_c$ and the decrease in susceptibility are explained by the decrease in the DOS due to the pseudogap formation. However, in Bi-2223, $T^*_{\chi}$ does not coincide with $T^*_{\rho_c}$. This is also considered to be result of inequivalent hole doping between the inner and outer planes. The susceptibility of Bi-2223 is considered to be the sum of the susceptibilities of inner plane and outer planes. Then, the negative temperature dependence in outer planes may conceal the pseudogap effect of the inner plane in the overdoped region, because the overall magnitude and the weight of the outer planes are larger than those of the inner plane.

On the other hand, $T^*_{\rho_{ab}}$ is also considered to be an indication of pseudogap formation~\cite{wata1,itoPRL}, because, in the case of YBa$_2$Cu$_3$O$_{6+\delta}$, $T^*_{\rho_{ab}}$ coincides with $T^*_{\rho_{c}}$~\cite{takenaka} as well as $T^*_\chi$. However, in Bi$_2$Sr$_2$CaCu$_2$O$_{8+\delta}$ (Bi-2212), we have previously pointed out that $T^*_{\rho_{ab}}$ does not coincide with $T^*_{\rho_{c}}$. The same tendency can be seen in the case of Bi-2223. The anomaly seen at $T^*_{\rho_{ab}}$ can be understood if we consider the strongly $k$-dependent quasiparticle lifetime. The carriers around the hot spot, where the pseudogap first opens up, will not contribute to the in-plane conduction, in contrast to $\rho_c$.

In summary, we measured a doping dependence of the thermopower, in-plane resistivity, out-of-plane resistivity, and susceptibility of high quality single crystal Bi-2223 for the first time. The room temperature thermopower, as well as the absolute value of resistivity ($\rho_{ab}$ and $\rho_c$), and c-axis length, continuously decrease with increasing $\delta$, and the overall magnitude of susceptibility increases with increasing $\delta$. All the results indicate that the carrier is properly controlled by our annealing method in the whole doping region. When the doping proceeds from underdope to optimum-dope, transition temperature $T_c$ increases quite similar to Bi-2212. However, it does not change in the overdoped region. On the other hand, we clearly observed the pseudogap formation in the $\rho_{ab}$ and $\rho_c$, which is similarly seen in Bi-2212. From the doping dependence of $\rho_c$, which is very sensitive to the onset of the pseudogap formation, we found that the pseudogap formation temperature $T^*$ does not change from its optimum value in the overdoped region. These results suggest that there is large difference in the carrier concentration between the inner and outer planes and in the overdoped region, the carriers may be mostly doped in the outer planes and the inner plane would remain at the underdoped level. As pointed by Kivelson, the combination of a large superconducting gap $\Delta$ in the inner plane and a large superfluid density of the outer planes would help to keep the highest $T_c$ in the overdoped region. On the other hand, $T^*_{\rho_c}$ pinning indicates that the pseudogapped inner plane also determines the $c$-axis transport property. This suggests that the carriers in the normal state are weakly coupled or confined to individual CuO$_2$ planes.


\end{document}